\documentclass[twoside,a4paper,11pt]{proceedings}
\usepackage{graphicx}
\usepackage{hyperref}
\usepackage{movie15}
\usepackage{natbib}

\begin{document}
\pagenumbering{arabic}
\pagestyle{myheadings}
\thispagestyle{empty}
\vspace*{-1cm}
{\flushleft\includegraphics[width=8cm,viewport=0 -30 200 -20]{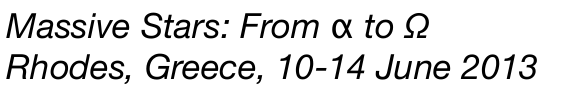}}
\vspace*{-0.4cm}
\begin{flushleft}
{\bf {\LARGE
The role of low-mass star clusters in massive star formation in Orion
}\\

\vspace{0.9cm}
V. M. Rivilla$^1$,
J. Mart\'in-Pintado$^2$,
I. Jim\'enez-Serra$^2$,
and
A. Rodr\'iguez-Franco$^1$
%
}\\
\vspace*{0.5cm}
%
$^{1}$
Centro de Astrobiolog\'ia (CSIC-INTA), Ctra. de Torrej\'on-Ajalvir, km. 4, E-28850 Torrej\'on de Ardoz, Madrid, Spain \\
$^{2}$
European Southern Observatory, Karl-Schwarzschild-Str. 2, 85748, Garching, Germany \\

%
\end{flushleft}
\markboth{
The role of low-mass star clusters in massive star formation in Orion
}{
Rivilla et al.
}
\thispagestyle{empty}
\vspace*{0.4cm}
\begin{minipage}[l]{0.09\textwidth}
\ 
\end{minipage}
\begin{minipage}[r]{0.9\textwidth}
\vspace{0cm}
\section*{Abstract}{\small
To distinguish between the different theories proposed to explain massive star formation, it is crucial to establish the distribution, the extinction, and the density of low-mass stars in massive star-forming regions. We analyzed deep X-ray observations of the Orion massive star-forming region using the Chandra Orion Ultradeep Project (COUP) catalog. We found that pre-main sequence (PMS) low-mass stars cluster toward the three massive star-forming regions: the Trapezium Cluster (TC), the Orion Hot Core (OHC), and OMC1-S. We derived low-mass stellar densities of 10$^{5}$ stars pc$^{-3}$ in the TC and OMC1-S, and of 10$^{6}$ stars pc$^{-3}$ in the OHC. The close association between the low-mass star clusters with massive star cradles supports the role of these clusters in the formation of massive stars. The X-ray observations show for the first time in the TC that low-mass stars with intermediate extinction are clustered toward the position of the most massive star, which is surrounded by a ring of non-extincted low-mass stars. 
Our analysis suggests that at least two basic ingredients are needed in massive star formation: the presence of dense gas and a cluster of low-mass stars. The scenario that better explains our findings assumes high fragmentation in the parental core, accretion at subcore scales that forms a low-mass stellar cluster, and subsequent competitive accretion. 
\vspace{10mm}
\normalsize}
\end{minipage}


\vspace{-0.75cm}
\section{Introduction}

\vspace{-0.25cm}

The formation of massive stars (M$>$8M$_{\odot}$) is still not fully understood. Two competing theories have been proposed : i) monolithic gravitational collapse and core accretion (\citealt{mckee03}); and iii) competitive accretion of low- and intermediate-mass stars in the cluster potential well (\citealt{bonnell06}). 
While the classic theory of monolithic core accretion predicts the formation of
mainly massive companions, competitive accretion models require a dense population of low-mass stars that form a cluster. Therefore, to distinguish between them it is crucial to establish the distribution and density of low-mass stars in massive star cradles. 
In this work we study the morphology and clustering of PMS low-mass stars in Orion, the nearest (d$\sim$414 pc) region of high-mass star formation, using the X-ray source catalog provided by the Chandra Orion Ultradeep Project (COUP; \citealt{getman05a}). 




\vspace{-0.25cm}
\section{Results}

\vspace{-0.25cm}

We analyzed the distribution of PMS low-mass stars with X-ray emission in Orion observed by Chandra as a function of extinction, with two different methods: a spatial gridding and a close-neighbors method, with cells of $\sim$0.03$\times$0.03 pc$^{2}$, the typical size of protostellar cores. Our results (Fig. \ref{Fig2}; see also Fig. 1 from \citealt{rivilla13a}) show that PMS low-mass stars cluster toward the three regions related to massive stars: the Trapezium Cluster (TC), the Orion Hot Core (OHC), and the OMC1-S region. We derived PMS low-mass stellar densities of 10$^{5}$ stars pc$^{-3}$ in the TC and OMC1-S, and 10$^{6}$ stars pc$^{-3}$ in the OHC. 
The morphology of the whole low-mass stellar cluster (see Fig. 1 from \citealt{rivilla13a}) is very similar to the results of the competitive accretion simulations by  \citet{bonnell04} (see their Fig. 5 and also Fig. 14 from \citealt{zinnecker07}), with massive stars located in the center of dense subclusters of low-mass stars.

The X-ray observations show in the TC that low-mass stars with intermediate extinction are clustered toward the position of the most massive star $\theta^{1}$ Ori C, which is surrounded by a ring of non-extincted PMS low-mass stars (Fig. \ref{Fig2}). This 'envelope-core' structure, also supported from infrared and optical observations (\citealt{rivilla13a}), is a natural outcome of the competitive accretion process. The non-extincted stars in the envelope of the TC could have lost the battle of competitive accretion - keeping their low-mass - against the four Trapezium stars in the core, which would have grown until they reached their current high masses. 
The non-extincted PMS stars in the envelope of the TC are therefore devoid of gas while those in the core are still partially embedded in gas and dust. 


{\small
\begin{figure}
\vspace{-2.5cm}
\hspace{-0.4cm}
\includegraphics[width=16.5cm]{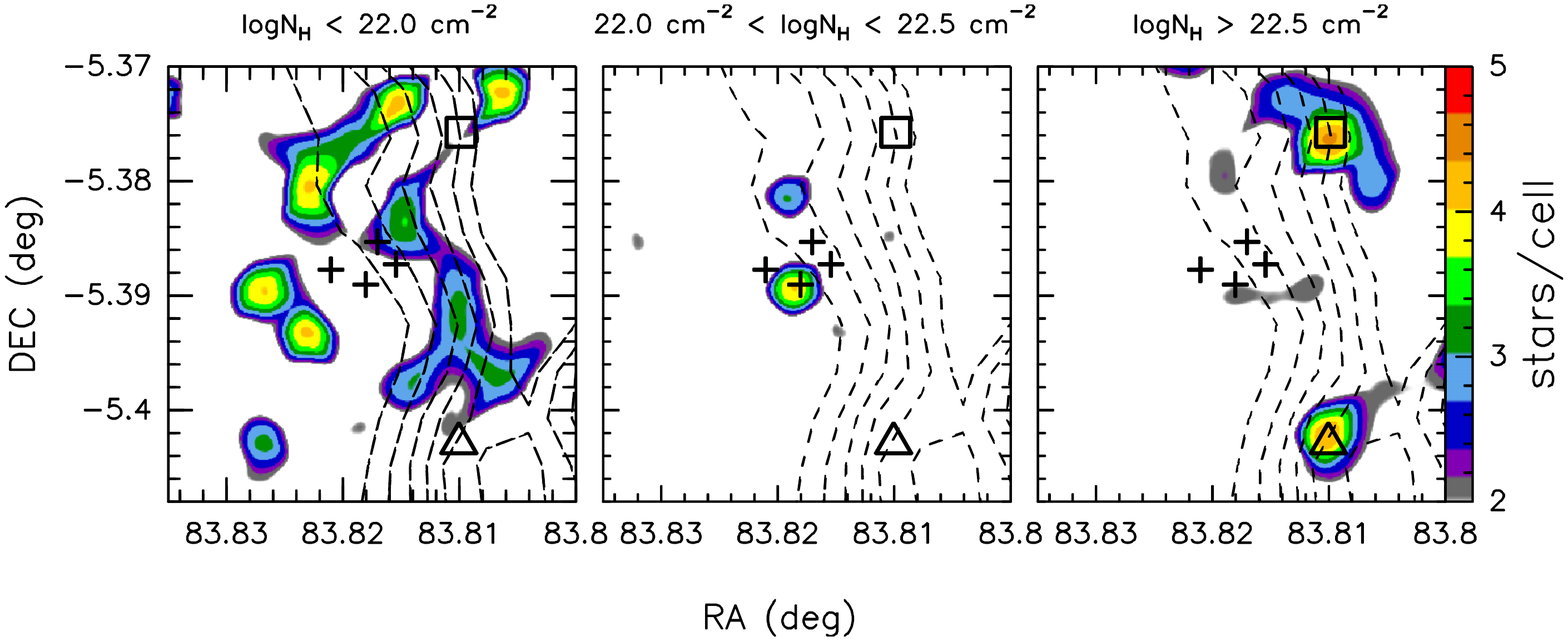} 
\vspace{-2.3cm}
\caption{\small Stellar density of low-mass stars for three different extinction ranges: log$N_{\rm H}\textless$22.0 cm$^{-2}$ (left panel), 22.0 cm$^{-2}\textless$log$N_{\rm H}\textless$22.5 cm$^{-2}$ (middle panel), and log$N_{\rm H}\textgreater$22.5 cm$^{-2}$ (right panel). We counted the number of COUP sources using a cell size of 15$^{\prime\prime}$ $\times$ 15$^{\prime\prime}$ (0.03 pc $\times$ 0.03 pc) (right color scale). Dashed contours represent dense gas from the Orion Molecular Cloud traced by the integrated intensity emission of CN (N=1-0) (\citealt{rodriguez-franco98}). Four crosses, the open square, and the open triangle show the location of the four main sequence massive Trapezium stars, the OHC, and the OMC1-S.}      
\label{Fig2}
\end{figure}
}

\vspace{-0.25cm}
\section{Discussion}

\vspace{-0.25cm}

We propose several scenarios for massive star formation in which the theories based on accretion (core accretion and competitive accretion) play different roles and the parental core presents different levels of fragmentation (see the sketch presented in Fig. 7 from \citealt{rivilla13a}): i) low fragmentation and monolithic core accretion; ii) intermediate fragmentation and subcore accretion; iii) high fragmentation, subcore accretion and subsequent competitive accretion. 

The first scenario, based on the classic core accretion theory (core-to-star), does not explain the formation of a significant population of low-mass stars within the cores, contradicting the results from the Chandra observations. 

The second scenario is also not fully appropriate to explain our results. In this scenario fragmentation does not always produce more massive subcores at the center of the parental core. Therefore, we would not expect the trend for massive stars to form at the center of the cluster observed in the TC (\citealt{allison09a}). Moreover, the presence of a significant low-mass stellar population revealed by X-rays in the massive star cradles suggests that the level of fragmentation of the initial core is high. 

This lead us to a third scenario, with higher fragmentation forming a low-mass stellar cluster via subcore accretion. These $^{\prime\prime}$stellar seeds$^{\prime\prime}$ can gather enough mass by subsequent competitive accretion, forming more massive stars at the cluster center, as observed in the TC . This process would also explain the structure of extincted core and non-extincted envelope found in the TC.  




\section{Conclusions}

\vspace{-0.25cm}

We found that PMS low-mass stars cluster toward the three massive star-forming regions in Orion: the Trapezium Cluster (TC), the Orion Hot Core (OHC), and the OMC1-S region. The close association between the highly-dense low-mass star clusters with massive star cradles supports the role of these clusters in the formation of massive stars. The X-ray observations show that PMS low-mass stars with intermediate extinction in the TC are clustered toward the position of the most massive star $\theta^{1}$ Ori C, which is surrounded by an envelope of non-extincted low-mass stars. Our analysis suggests that not only dense gas is needed to form massive objects, but also a cluster of low-mass stars. The scenario that better explains our findings implies: i) high fragmentation in the parental core; ii) accretion at subcore scales forming a low-mass stellar cluster; and iii) subsequent competitive accretion that form massive stars at the cluster center.

\small  
%
%


{\small
\bibliographystyle{mn2e}
\bibliography{proceedings}
}

\end{document}